\documentclass{article}
\usepackage{etex}
\reserveinserts{28}
\usepackage{setspace, amsmath, amssymb, amsthm}
\usepackage[colorlinks = true]{hyperref}
\usepackage[margin = 2.5cm]{geometry}
\usepackage[english]{babel}
\usepackage[utf8x]{inputenc}
\usepackage{amsfonts,tikz}
\usepackage{graphicx} 
\usepackage{epsfig}
\usepackage{epstopdf}
\usepackage{subcaption}
\usetikzlibrary{calc}
\usepackage{csquotes}
\usepackage{caption}
\usepackage{array}
\usepackage{tkz-graph}
\usetikzlibrary{fit}
\usepackage{tikz-cd}
\usetikzlibrary{arrows}
\usetikzlibrary{positioning}
\newtheorem{theorem}{Theorem}[section]
\newtheorem{corollary}{Corollary}[theorem]

\newtheorem{mydef-remark}{Remark}
\tikzstyle{vertex}=[circle, draw, inner sep=0pt, minimum size=4pt]
\label{marker}

\newcommand{\ket}[1]{|{#1}\rangle}
\newcommand{\bra}[1]{\langle{#1}|}
\newcommand{\abs}[1]{\lvert  #1  \rvert}
\newcommand{\card}[1]{\lvert \lvert  #1 \lvert \rvert}

\begin{document}
\title{Separability of Graph Laplacian Quantum States: Utilizing Unitary Operators, Neighbourhood Sets and Equivalence Relation }
 \author{Anoopa Joshi\\Indian Institute of Technology Jodhpur, Rajasthan, India\\Parvinder Singh \\Central University of Punjab, Bathinda, India\\Atul Kumar \\ Indian Institute of Technology Jodhpur, Rajasthan, India \thanks{
Email: \texttt{atulk@iitj.ac.in}} }
 
\date{}

\maketitle
\begin{abstract}
	This article delves into an analysis of the intrinsic entanglement and separability feature in quantum states as depicted by graph Laplacian. We show that the presence or absence of edges in the graph plays a pivotal role in defining the entanglement or separability of these states. We propose a set of criteria for ascertaining the separability of quantum states comprising $n$-qubit within a composite Hilbert space, indicated as $H=H_1 \otimes H_2 \otimes \dots \otimes H_n$. This determination is achieved through a combination of unitary operators, neighbourhood sets, and equivalence relations.
 
\end{abstract}

\section{Introduction} 
In classical and quantum mechanics, a pure state can be described through a point in phase space and a vector complex vector space, respectively. Although pure states represent states with maximal information, mixed states are characterized by a degree of uncertainty due to the uncertainty principle. The 
superposition principle leads to one of the most notable aspects of quantum mechanics- entanglement - which
evidently differentiates it from classical mechanics. Clearly, entanglement is a fundamental phenomenon of quantum mechanics \cite{einstein1935can,hall2013quantum,peres2006quantum,schrodinger1983naturwissenschaften} and has been established as an important resource for quantum information theory \cite{berkolaiko2017elementary,braunstein2006laplacian,dur2003multiparticle}. Some of the key application of entanglement are quantum computation \cite{gruska1999quantum,nielsen2002quantum}, quantum dense coding \cite{bennett1992communication,gisin2002quantum,gruska1999quantum,nielsen2002quantum}, quantum teleportation \cite{bennett1992communication,bennett1993teleporting}, quantum secret sharing protocols \cite{hillery1999quantum}, and quantum key distribution protocols \cite{ekert1991quantum,ekert1992quantum}. For 
two-qubit systems, classification and quantification of quantum entanglement is well established. For example, the necessary and sufficient condition for separability of two-qubit systems has been given by Peres \cite{peres1996separability,peres1996chaotic,peres1997quantum,peres2006quantum} and Horodecki \textit{et al.} \cite{horodecki1996separability,horodecki1996teleportation,horodecki2001separability,horodecki2009quantum}. However, transitioning from the analysis of entanglement in two-qubit states to extending the study to $n$-qubit sates poses a formidable challenge. Since entanglement cannot be shared freely among many particles, entanglement versus separability for $n$-qubit states is a complex phenomenon. Graph theory \cite{bapat2010graphs,bapat2014adjacency,west2001introduction} provides a way forward with a physical interpretation in various physical and chemical theories. In fact, the graph theoretical approach has made significant contributions to quantum physics \cite{hall2013quantum} and information theory \cite{berkolaiko2017elementary,braunstein2006laplacian,braunstein2006some,dur2003multiparticle}. Graph theory leads to a pictorial representation of quantum states \cite{adhikari2017laplacian,braunstein2006laplacian,braunstein2006some,dur2003multiparticle,dutta2016bipartite,hassan2007combinatorial,joshi2018concurrence,joshi2020chaotic,joshi2022entanglement}, called graph Laplacian quantum state. In quantum mechanics, a density operator is a positive semi-definite, Hermitian matrix with unit trace, acting on a Hilbert space, say $H(A)$, and density operator have graphical interpretation.

The density operator, representing a state associated with a graph, corresponds to the normalized Laplacian matrix. Given a simple graph $G = (V(G), E(G))$ with vertex set $V(G)$ and edge set $E(G)$, the Laplacian matrix of $G$, denoted as $L(G)$, is singular and positive semi-definite, and can be expressed as $L(G)=D(G)-A(G)$, where $D(G)$ is the diagonal matrix of vertex degrees, and $A(G)$ is the adjacency matrix. Throughout this paper, we concentrate on studying quantum states associated with graphs.
\subsection{Preliminaries}
In this section, we briefly describe the basic terminologies used in this article to facilitate the discussions on our analysis and results obtained therein. 
\newline
\textbf{\textit{Graph}}: A graph, denoted as $G = (V(G), E(G))$, comprises a set of vertices $V(G)=\{v_{i} \}$ and an edge set $E(G)=\{e_{ij} \}$. Each edge weight from vertex $v_i$ to vertex $v_j$ is represented as $a_{ij}$. The basis vector set $B = \{ b_{i} \}$ consists of column vectors associated with the graph $G$. The Laplacian matrix of the graph $G$ is denoted as $L(G)$. For a weighted graph on the real numbers $\mathbb{R}$, the Laplacian matrix $L(G)$ is defined as $ L(G)= D(G) - A(G) + D_0(G)$, where $D(G) = [d_{ij}]$ and $D_0(G)= [d_{0_{ij}}]$. The elements $d_{ij}$ and $d_{0_{ij}}$ are specified as 
 $	d_{ij} = \begin{cases}
			\sum a_{il}, \text{ for } i=j \\
			0, \text{ otherwise }
		\end{cases} $   and 
  $ d_{0_{ij}} = \begin{cases}
			 \text{ loop weight on vertex $v_i$}  \text{ for } i=j \\
			0, \text{ otherwise }
		\end{cases}$, respectively. For a weighted graph on the complex numbers $\mathbb{C}$, the Laplacian matrix $L$ is defined as $L(G) = D(G) + A(G) - D_0(G)$, where the elements $d_{ij}$ and $d_{0_{ij}}$ are given as 
 $d_{ij}=\begin{cases}
	\sum_{v_i \in V(G),v_i \neq v_j} \abs{a_{ij}}+a_{ii}, \text{ for } i=j \\
	0, \text{ otherwise }
\end{cases}$ and 
$d_{0_{ij}}= \begin{cases}  \text{ loop weight on vertex $v_i$}  \text{ for } i=j \\
	0, \text{ otherwise }
\end{cases}$, respectively. 

Since the density operator $(\rho)$ of a quantum state is described by a Hermitian positive semi-definite matrix with unit trace \cite{barnett2009quantum},  $\rho$ corresponding to a graph $G$ is defined by a normalised Laplacian matrix \cite{hassan2007combinatorial,braunstein2006laplacian,joshi2018concurrence,joshi2022entanglement} 
$$ \rho_G=  \frac{1}{Tr(L(G))}[L(G)].$$

Here in term of basis, $L(G)=  [ L_{ij} ] = [ l_{ij} {b_i}^T b_j]$,  where  $V=\{ b_i={c_1}^i\otimes {c_2}^i \otimes \dots \otimes {c_n}^i \text{ for all $i= 1,2,\dots ,2^n$}\}$ be the vertex set and $b_i$ is column basis.

\textbf{\textit{Neighbourhood set of a vertex}}: The set of neighbour of a vertex of $v_i$ is defined as $ N(v_i)= \{v_j : (v_i,v_j) \in E(G) \} \cup {v_i}$, thus for a simple graph $N(v_i)=d(v_i)$. If $v_i$ is an isolated vertex then $v_i$ has only one neighbour itself \cite{sampatkumar1985neighbourhood}.

\textbf{\textit{Equivalence relation}}:  An equivalence relation is a relationship on a set. Let $N$ be a set and $"\sim"$ be a relation on $N$. The relation $"\sim"$  is an equivalence relation for all $a,b,c \in N $ if it is \begin{enumerate}
	\item Reflexive: $a \sim a$  \item Symmetric:  if $a \sim b$ then $b \sim  a$  , and \item Transitive: if $a \sim b$ and $b \sim c$ then $a \sim c$.
\end{enumerate}

\textbf{\textit{ Partial quantum gate on a graph $G$}}:
For this, we first define a unitary operator $U$ with combination of $I_2$ and $\sigma_x$ as $U= \underbrace{ I_2 \otimes \dots \otimes I_2}_{q} \otimes {U_{q+1} \otimes \dots \otimes U_{q+p} }$, where $U_{q+1} , \dots, U_{q+p}  $ are unitary operators represented either by $I_2$ or $\sigma_x$. If $G$ is a graph on $2^n$ vertices and $V=\{ b_i={c_1}^i\otimes {c_2}^i \otimes \dots \otimes {c_n}^i \text{ for all $i= 1,2,\dots ,2^n$}\}$ is the vertex set where $c_i's$ are  column basis vector in ${\mathbb{C}}^2$, then the partial quantum gate on the graph is defined as $(\text{Partial}U)G= (Ub_i,Ub_j) \in E(G_1)$ for all $(b_i,b_j) \in E(G)$, where $U_k= \begin{cases}
I_2  \text{ if } {c_k}^i= {c_k}^j\\
\sigma_x \text{ if } {c_k}^i \neq {c_k}^j
\end{cases}$ 
   
\textbf{\textit{Partial quantum gate on density operator}}: If  $\rho_G= \frac{1}{L(G)} \big[  e_{ij} b_i \otimes {b_j}^T \big]_{2^n \times 2^n}$ is a density operator, which is a block matrix with dimensions $2^q \times 2^q $, of the graph $G$ on $2^{n=p+q}$ vertices and $V=\{ b_i={c_1}^i\otimes {c_2}^i \otimes \dots \otimes {c_n}^i \text{ for all $i= 1,2,\dots ,2^n$}\}$ is the vertex set where $c_i's$ are  column basis vector in ${\mathbb{C}}^2$, then the partial quantum gate on the density operator is defined as $(\text{Partial}U) \rho_G= \frac{1}{L(G)} \big[  e_{ij} Ub_i \otimes {(Ub_j)}^T \big]_{2^n \times 2^n}$,
where $U_k= \begin{cases}
I_2  \text{ if } {c_k}^i= {c_k}^j\\
\sigma_x \text{ if } {c_k}^i \neq {c_k}^j
\end{cases}$

\textit{As per the standard definition:} \begin{enumerate} \item $(G,a)$ is a graph associated with an $n$-qubit quantum state, having $2^{p+q}$ vertices labeled by $(ij)$, $i = 1, \dots , 2^p$ and $j = 1, \dots, 2^q$.
    \item For a given matrix $A$, $\overline{A}$ denotes the complex conjugate of each element of matrix $A$ and $\card{E(G}$ represents cardinality of edge set or number of edges.
\end{enumerate}
 \section{Entanglement and Separability of a Graph Associated n-Qubit State}
 Let $G$ be a graph on $2^n$ vertices associated with an $n$-qubit state and $\rho_G$ be the density operator of the graph $G$. The graph $G$ is known to be associated with a separable state if $\rho_G$ can be represented as a tensor product of two or more density operators, which correspond to graphs of order $2^m$ ($m < n$). In this section, we address the problem of quantum separability in graphs associated with quantum states by applying  quantum gates (combination of $I_2$ and $\sigma_x$) on the density operator. The issue of quantum inseparability of mixed states has garnered significant attention from researchers and has been explored in various physical contexts \cite{petz2007quantum}. Werner (1989) introduced the concept of separable states, where the identification of a quantum state as separable or inseparable is known as the quantum separability problem. Specifically, a bipartite state is considered separable if and only if it can be expressed as a convex combination of product states. For an $n$-partite state on a Hilbert space $H=H_1 \otimes H_2 \otimes H_3 \otimes \dots \otimes H_n$, separability is characterized by the ability to represent the state as a convex sum of tensor products of subsystem states.

 \begin{theorem}  \label{TH1}
 Let the density operator $ \rho_G = [A^{xy}]_{2^q \times 2^q} $ be a block matrix of a weighted graph $G$ on $2^{(n=p+q)}$ vertices. The graph $G$ will be associated with a separable state if and only if $A^{xy}=B^{xy} - C^{xy}+ iD^{xy}-iE^{xy} $ for all $x \neq y$, and $A^{xx}- \sum_{x \neq y}B^{xy}- \sum_{x \neq y}C^{xy} - \sum_{x \neq y}D^{xy}-\sum_{x \neq y}E^{xy}  \geq 0$ for all $x$, where  $B^{xy}$, $C^{xy}$, $D^{xy}$ and $E^{xy}$ are positive semi-definite matrices.   
\end{theorem}
Proof $\rho_G = [A^{xy}]_{2^q \times 2^q} $ is a block matrix of a weighted graph $G$ on $2^{(n=p+q)}$ vertices. We know that a composite system is said to be separable if the density operator $\rho_G$ can be written as a convex combination of product states, i.e.,
\begin{equation} \label{eq1}
\rho_G =\sum_{k} p_k {\rho_1}^k \otimes {\rho_2}^k,
\end{equation}	
where $p_k \geq 0$, $\sum_{k}p_k =1$, and ${\rho_1}^k$ and ${\rho_2}^k$  are density operators of individual subsystems. From equation (\ref{eq1}), we can say that blocks $A^{xy}$ can be decomposed as $A^{xy}=B^{xy} - C^{xy}+ iD^{xy}-iE^{xy} $ for all $x \neq y$, where $B^{xy}$, $C^{xy}$, $D^{xy}$ and $E^{xy}$ are positive semi-definite matrices, and $A^{xx}- \sum_{y \neq x}B^{xy}- \sum_{y \neq x}C^{xy} - \sum_{y\neq x}D^{xy} -\sum_{y \neq x }E^{xy}   \geq 0$ for all $x$. \\
 Conversely, if $A^{xy}=B^{xy} - C^{xy}+ iD^{xy}-iE^{xy} $ for all $x \neq y$, where $B^{xy}$, $C^{xy}$, $D^{xy}$ and $E^{xy}$ are positive semi-definite then we have $A^{xx}- \sum_{y \neq x}B^{xy}- \sum_{y \neq x}C^{xy} - \sum_{y\neq x}D^{xy} -\sum_{y \neq x }E^{xy}   \geq 0$ for all $x$, and therefore the density operator can be written as a  convex combination of product states \cite{joshi2022entanglement}. 

\textbf{Example:}  The graphs $G_1$, $G_2$, and $G_3$ below serve as illustrations of the aforementioned theorem.
 
 $$	\begin{tikzpicture}[auto, node distance=4cm, every loop/.style={},
thick,main node/.style={circle,draw,font=\sffamily\Large\bfseries}]
\node[main node] (1) {$\ket{00}$};
\node[main node] (2) [ right of=1] {$\ket{01}$};
\node[main node] (3) [ below of=1] {$\ket{10}$};
\node[main node] (4) [ below of=2] {$\ket{11}$};
\draw (2, -6) node[below] { $ \hspace{1cm} G_1 \hspace{1cm}$};   	    	 
\path[every node/.style={font=\sffamily\small}]
(1) edge node  [] {-2-i} (2)
(1) edge node  [] {1} (3)
(1) edge node  [left] {i} (4)
(1) edge [loop above] node[]{1- $\sqrt{5}$} (1)
(2) edge node  [right] {-2+i} (3)
(2) edge node  [] {1} (4)	
(2) edge [loop above] node[]{1- $\sqrt{5}$} (2)
(3)edge node  [] {-2-i} (4) 
(3) edge [loop below] node[]{2- $2\sqrt{5}$} (3)
(4) edge [loop below] node[]{2- $2\sqrt{5}$} (4);
\end{tikzpicture} $$
				
$\rho_{G_1} = \frac{1}{12} \begin{bmatrix} 3 & -2-i & 1 & i \\ -2+i & 3 & -2+i & 1 \\ 
1 & -2-i & 3 & -2-i \\ -i & 1 & -2+i & 3 \end{bmatrix}
= \begin{bmatrix} A & B\\B^{\dagger} & C \end{bmatrix} 
= \frac{1}{3} \Big\{\frac{1}{2}\begin{bmatrix} 1&i\\-i&1 \end{bmatrix} 
\otimes \frac{1}{2}  \begin{bmatrix} 1 & -i\\i & 1 \end{bmatrix} \Big\} \\
+ \frac{1}{3} \Big\{\frac{1}{2}  \begin{bmatrix} 1 & -i\\i & 1 \end{bmatrix}
\otimes \frac{1}{2}  \begin{bmatrix} 1 & -1\\ -1 & 1 \end{bmatrix} \Big\}+
\frac{1}{3} \Big\{\frac{1}{2}  \begin{bmatrix} 1 & 1\\1 & 1 \end{bmatrix} 
\otimes \frac{1}{2}  \begin{bmatrix} 	1 & -1\\-1 & 1 \end{bmatrix}\Big\} ,$ 

where $B=  \begin{bmatrix}	1 & i\\-2+i & 1 \end{bmatrix} 
= \begin{bmatrix}	1 & -1\\-1 & 1 \end{bmatrix}-i\begin{bmatrix} 1 & -1\\-1 & 1 \end{bmatrix} +
i \begin{bmatrix} 1& -i\\ i&1  \end{bmatrix}   $, \\
 and $ \begin{bmatrix}	 3 & -2-i  \\ -2+i & 3 	 \end{bmatrix}			 
 - \begin{bmatrix}	1 & -1\\-1 & 1	\end{bmatrix}-\begin{bmatrix} 1 & -1\\-1 & 1 \end{bmatrix}
 - \begin{bmatrix}	1 & -i\\-i & 1	\end{bmatrix} = \begin{bmatrix} 0 & 0\\0 & 0	\end{bmatrix}$.

$$	\begin{tikzpicture}[auto, node distance=4cm, every loop/.style={},
    thick,main node/.style={circle,draw,font=\sffamily\Large\bfseries}]
   \node[main node] (1) {$\ket{00}$};
  \node[main node] (2) [ right of=1] {$\ket{01}$};
  \node[main node] (3) [ below of=1] {$\ket{10}$};
  \node[main node] (4) [ below of=2] {$\ket{11}$};
  \draw (2, -6) node[below] { $ \hspace{1cm} G_2 \hspace{1cm}$};   	    	 
 \path[every node/.style={font=\sffamily\small}]
 (1) edge node  [] {1} (2)
 (1) edge node  [] {i} (3)
 (1) edge node  [left] {i} (4)
 (1) edge [loop above] node[]{-2} (1)
 (2) edge node  [right] {i} (3)
 (2) edge node  [] {i} (4)	
 (2) edge [loop above] node[]{-2} (2)
 (3)edge node  [] {1} (4) 
 (3) edge [loop below] node[]{-2} (3)
 (4) edge [loop below] node[]{-2} (4);
 \end{tikzpicture} $$
				
 $\rho_{G_2} = \frac{1}{4} \begin{bmatrix}  1 & 1 & i & i\\ 1 & 1 & i & i\\  -i & -i & 1 & 1\\ -i & -i & 1 & 1\\ \end{bmatrix} = \begin{bmatrix}	A & B\\B^{\dagger} & C \end{bmatrix} 
  =  \frac{1}{2}\begin{bmatrix}  1&i\\-i&1 \end{bmatrix} 
  \otimes \frac{1}{2}  \begin{bmatrix} 1 & 1\\1 & 1 \end{bmatrix} $,\\
  where $B=  \begin{bmatrix} 	i & i\\ i & i \end{bmatrix} = i\begin{bmatrix} 	1 & 1\\1 & 1  \end{bmatrix}$.  

We further discuss the separability of 3-qubit state $\psi= \frac{1}{\sqrt{2}} ( \ket{0} \otimes \ket{1} \otimes \ket{+} - \ket{1} \otimes \ket{0} \otimes \ket{-})$ where the graph $G_3$ is associated with the state $\psi$. Therefore, the density operator $$\rho_{G_3}= \frac{1}{4}\begin{bmatrix}
       0 & 0 & 0 & 0 & 0 & 0 & 0 & 0 \\ 
 0 & 0 & 0 & 0 & 0 & 0 & 0 & 0 \\  0 & 0 & 1 & 1 & -1 & -1 & 0 & 0\\
 0 & 0 & 1 & 1 & -1 & -1 & 0 & 0 \\ 0 & 0 & -1 & -1 & 1 & 1 & 0 & 0 \\ 
 0 & 0 & -1 & -1 & 1 & 1 & 0 & 0 \\ 0 & 0 & 0 & 0 & 0 & 0 & 0 & 0 \\ 
 0 & 0 & 0 & 0 & 0 & 0 & 0 & 0 \end{bmatrix}$$

$$	\begin{tikzpicture}[auto, node distance=2.5cm, every loop/.style={},
thick,main node/.style={circle,draw,font=\sffamily\Large\bfseries}]
\node[main node] (1) {$\ket{000}$};	\node[main node] (2) [ right of=1] {$\ket{001}$};
\node[main node] (3) [ right of=2] {$\ket{010}$};
\node[main node] (4) [ right of=3] {$\ket{011}$};
\node[main node] (5) [ below of=1] {$\ket{100}$};
\node[main node] (6) [ below of=2] {$\ket{101}$};
\node[main node] (7) [ below of=3] {$\ket{110}$};
\node[main node] (8) [ below of=4] {$\ket{111}$};
\draw (4,-4) node[below] { $ \hspace{1cm} G_3 \hspace{1cm}$};
\path[every node/.style={font=\sffamily\small}]
(3) edge [] node[] {-1} (4)
(3) edge [] node[] {1} (5)
(3) edge [] node[] {1} (6)
(4)edge [] node[] {1} (5)
(4) edge [] node[] {1} (6)
(5) edge [] node[] {-1} (6);
\end{tikzpicture} $$ 
Clearly, we can re-express $\rho_{G_3}$ as 
$$\rho_{G_3}= \frac{1}{4}\begin{bmatrix}
       0 & 0 & 0 & 0 & 0 & 0 & 0 & 0 \\ 
 0 & 0 & 0 & 0 & 0 & 0 & 0 & 0 \\  0 & 0 & 1 & 1 & -1 & -1 & 0 & 0\\
 0 & 0 & 1 & 1 & -1 & -1 & 0 & 0 \\ 0 & 0 & -1 & -1 & 1 & 1 & 0 & 0 \\ 
 0 & 0 & -1 & -1 & 1 & 1 & 0 & 0 \\ 0 & 0 & 0 & 0 & 0 & 0 & 0 & 0 \\ 
 0 & 0 & 0 & 0 & 0 & 0 & 0 & 0 \end{bmatrix} = 
 \begin{bmatrix} A_{11} & A^{12} & A^{13} & A^{14} \\
A_{21} & A^{22} & A^{23} = -  B^{23}  & A^{24} \\ A_{31} & A^{32} = - B^{32} & A^{33} & A^{34} \\
A_{41} & A^{42} & A^{43} & A^{44} \end{bmatrix}$$\\
Here we can see that $ A^{22} - B^{23} \geq 0 $ and $ A^{33} - B^{32} \geq 0 $. Hence, the state is separable.

\begin{corollary} \label{C1}
    Let the density operator $ \rho_G = [A_{ij}]_{2^q \times 2^q} $ be a block matrix of a weighted graph $G$ on $2^{(n=p+q)}$ vertices where  $V=\{ b_i={c_1}^i\otimes {c_2}^i \otimes \dots \otimes {c_n}^i \text{ for all $i= 1,2,\dots ,2^n$}\}$ be the vertex set and $c_i's$ are  basis vectors in ${\mathbb{C}}^2$. Consider a  unitary operator $U= \underbrace{ I_2 \otimes \dots \otimes I_2}_{q} \otimes {U_{q+1} \otimes \dots \otimes U_{q+p} }$ such that $U_{q+1} , \dots, U_{q+p}  $ are either $I_2$ or $\sigma_x$. If $ (\overline{\text{Partial}}U) \rho_G=\rho_G $  and $A_{ii}- \sum_{j \neq i}B_{ij}- \sum_{j \neq i}C_{ij} \geq 0$ for all $i$, so that $U_k= \begin{cases} I_2  \text{ if } {c_k}^i= {c_k}^j\\ \sigma_x \text{ if } {c_k}^i \neq {c_k}^j \end{cases}$, and $A_{ij} =B_{ij} - C_{ij}$ $\forall i \neq j$ where $B_{ij}$ and $C_{ij}$ are positive semi definite matrices, then the graph $G$ is associated with a separable state. 
\end{corollary}

Proof: Let the density operator $ \rho_G = [A_{ij}]_{2^q \times 2^q} $ be a block matrix of a weighted graph $G$ on $2^{(n=p+q)}$ vertices where  $V=\{ b_i={c_1}^i\otimes {c_2}^i \otimes \dots \otimes {c_n}^i \text{ for all $i= 1,2,\dots ,2^n$}\}$ be the vertex set and $c_i's$ are   basis vectors in ${\mathbb{C}}^2$. The partial quantum gate on the density operator can be define as $(\text{Partial}U) \rho_G= \frac{1}{L(G)} [l_{ij} Ub_i \otimes {Ub_j}^T]$ where $U_k= \begin{cases} I_2 \text{ if } {c_k}^i= {c_k}^j\\\sigma_x \text{ if } {c_k}^i \neq {c_k}^j \end{cases}$. If $ (\overline{\text{Partial}}U)\rho_G=\rho_G$  then the blocks of density operator are Hermitian and can be decomposed as $A_{ij}= B_{ij}- C_{ij}$ $\forall i \neq j$ where $B_{ij}$ and $C_{ij}$ are positive semi definite matrices. Therefore, if $A_{ii}- \sum_{j \neq i}B_{ij}- \sum_{j \neq i}C_{ij} \geq 0$ for all $i$ then $\rho_G$ is decomposed as a convex sum of product states by theorem \ref{TH1}. Hence proved \cite{joshi2022entanglement}.

 \textbf{Example:} Let us consider the graph $G$ below such that
 $$ 	\begin{tikzpicture}[auto, node distance=4cm, every loop/.style={},
 thick,main node/.style={circle,draw,font=\sffamily\Large\bfseries}]
 \node[main node] (1) {$\ket{00}$};
 \node[main node] (2) [ right of=1] {$\ket{01}$};
 \node[main node] (3) [ below of=1] {$\ket{10}$};
 \node[main node] (4) [ below of=2] {$\ket{11}$};
 \draw (2, -5) node[below] { $ \hspace{1cm} G \hspace{1cm}$};   	 \path[every node/.style={font=\sffamily\small}]
 (1) edge node  [] {-4} (2)
 (1) edge node  [] {4} (3)
 (1) edge node  [left] {6} (4)
 (2) edge node  [right] {6} (3)
 (2) edge node  [] {4} (4)	
 (3)edge node  [] {-4} (4);
 \end{tikzpicture} $$
 
 $\rho_G = \frac{1}{24} \begin{bmatrix}6 & 4 &- 4 & -6\\4 & 6 & -6 & -4\\-4 & -6 & 6 & 4\\-6 & -4 & 4 & 6
\end{bmatrix} 
  = \begin{bmatrix}A & B\\B^T & C\end{bmatrix} $
  
  $= \frac{8}{24}\Big\{ \frac{1}{4}\begin{bmatrix}
 2& -2\\-2 & 2 \end{bmatrix} \otimes \frac{1}{2}  \begin{bmatrix} 1 & 1\\1 & 1 \end{bmatrix} \Big\} 
+ \frac{4}{24}\Big\{ \frac{1}{2}  \begin{bmatrix} 1 & 1\\1 & 1\end{bmatrix} 
\otimes \frac{1}{2}  \begin{bmatrix} 1 & -1\\ -1 & 1\end{bmatrix} \Big\} 
+ \frac{12}{24}\Big\{ \frac{1}{6}\begin{bmatrix} 3& -3\\-3 & 3 \end{bmatrix} 
\otimes \frac{1}{2}  \begin{bmatrix} 1 & 1\\1 & 1 \end{bmatrix} \Big\} ,$  

where $B=  \begin{bmatrix}	- 4 & -6\\-6 & -4 \end{bmatrix} 
= \begin{bmatrix} 1 & -1\\-1 & 1 \end{bmatrix}
-\begin{bmatrix} 3 & 3\\3& 3\end{bmatrix}
- \begin{bmatrix} 2& 2\\2 & 2 \end{bmatrix}   $, 

and $ \begin{bmatrix} 6 & 4\\4 & 6 \end{bmatrix} 
- \begin{bmatrix} 1 & -1\\-1 & 1 \end{bmatrix}
-\begin{bmatrix} 3 & 3\\3& 3 \end{bmatrix}
- \begin{bmatrix} 2& 2\\2 & 2 \end{bmatrix} 
= \begin{bmatrix} 0& 0\\0 & 0 \end{bmatrix}.$

 \begin{theorem} \label{TH3}
Let the density operator $ \rho_G = [A^{xy}]_{2^q \times 2^q} $ be a block matrix of a weighted graph $G$ on $2^{n=p+q}$ vertices where $A^{xy} = [a^{xy}_{ij}]_{2^p \times 2^p} $ and $V=\{ b_i={c_1}^i\otimes {c_2}^i \otimes \dots \otimes {c_n}^i \text{ for all $i= 1,2,\dots ,2^n$}\}$ be the vertex set such that $c_i's$ are  basis vectors in ${\mathbb{C}}^2$. 
 The graph $G$ is associated with a separable state if $A^{rs} A^{mn}= A^{mn} A^{rs}$.
\end{theorem}
    
 Proof Let $\rho_G= \frac{1}{L(G)} [l_{ij} b_i \otimes {b_j}^T] = [A^{xy}]_{2^q 2^q}$ $(A^{xy}= [a^{xy}_{ij}]_{2^p \times 2^p} )$ be a density operator of a weighted graph $G$ on $2^{n=p+q}$ vertices, and $V=\{ b_i={c_1}^i\otimes {c_2}^i \otimes \dots \otimes {c_n}^i \text{ for all $i= 1,2,\dots ,2^n$}\}$ be the vertex set. If $A^{rs} A^{mn}= A^{mn} A^{rs} $, then 
 $(\overline{\text{Partial}}U) \rho_G=\rho_G $ 
where $U= \underbrace{ I_2 \otimes \dots \otimes I_2}_{q} \otimes {U_{q+1} \otimes \dots \otimes U_{q+p} }$ such that $U_{q+1}, \dots, U_{q+p}  $ are either $I_2$ or $\sigma_x$ with 
$U_k= \begin{cases} I_2  \text{ if } {c_k}^i= {c_k}^j\\ \sigma_x \text{ if } {c_k}^i \neq {c_k}^j \end{cases}$. This further suggests that blocks $A^{xy}$ can be decomposed as $A^{xy}=B^{xy} - C^{xy}+ iD^{xy}-iE^{xy} $ for all $x \neq y$, where $B^{xy}$, $C^{xy}$, $D^{xy}$ and $E^{xy}$ are positive semi-definite matrices. 

If blocks commute, they also share the same set of eigenvectors. Let $\lambda_i$ denote the eigenvalues of $A^{xx}$, $\nu_i$ denote the eigenvalues of  $A^{xy}$, $\alpha_i$ denote the eigenvalues of  $B^{xy}$, $\beta_i$ denote the eigenvalues of  $C^{xy}$, $\gamma_i$ denote the eigenvalues of  $D^{xy}$, and $\delta_i$ denote the eigenvalues of  $E^{xy}$.
Since $B^{xy}$, $ C^{xy}$, $D^{xy}$, and $E^{xy}$ are positive semi-definite matrices, their eigenvalues are non-negative. Hence, $\lambda_i- (\alpha_i+\beta_i+\gamma_i+\delta_i) \geq0 $. Therefore, $ A^{xx} - \sum_{y \neq x} B^{xy} - \sum_{y \neq x} C^{xy} - \sum_{y \neq x}D^{xy} -\sum_{y\neq x}E^{xy}\geq 0 $, for all $x$ which further implies that density operator is associated with a separable state from Theorem (\ref{TH1}).

\textbf{Example:} The following graphs $G_1$ and $G_2$ represent illustrations of the above theorem.  
$$	
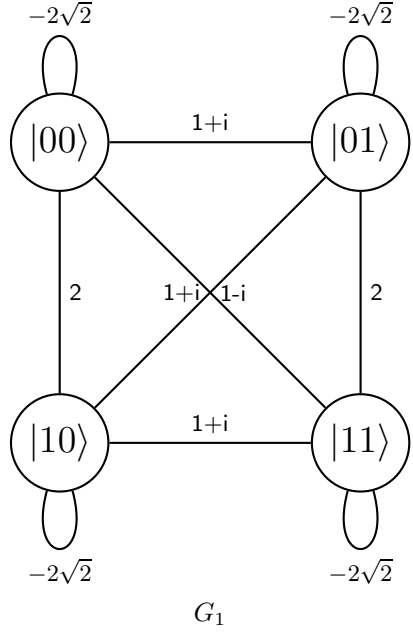
\begin{tikzpicture}[auto, node distance=4cm, every loop/.style={},
thick,main node/.style={circle,draw,font=\sffamily\Large\bfseries}]
\node[main node] (1) {$\ket{00}$};
\node[main node] (2) [ right of=1] {$\ket{01}$};
\node[main node] (3) [ below of=1] {$\ket{10}$};
\node[main node] (4) [ below of=2] {$\ket{11}$};
\draw (2, -6) node[below] { $ \hspace{1cm} G_1 \hspace{1cm}$};   	    	 
\path[every node/.style={font=\sffamily\small}]
(1) edge node  [] {1+i} (2)
(1) edge node  [] {2} (3)
(1) edge node  [left ] {1+i} (4)
(1) edge [loop above] node[]{ $-2\sqrt{2}$} (1)
(2) edge node  [right ] {1-i} (3)
(2) edge node  [] {2} (4)	
(2) edge [loop above] node[]{$-2\sqrt{2}$} (2)
(3)edge node  [] {1+i} (4) 
(3) edge [loop below] node[]{ $-2\sqrt{2}$} (3)
(4) edge [loop below] node[]{ $-2\sqrt{2}$} (4);		
\end{tikzpicture} $$
					
$\rho_{G_1} = \frac{1}{8} \begin{bmatrix} 2 & 1+i  & 2 & 1+i \\ 1-i & 2 & 1-i & 2 \\ 2 & 1+i & 2 & 1+i  \\ 1-i  & 2 & 1-i & 2 \end{bmatrix} = 
\begin{bmatrix}	A^{11} & A^{12}\\ A^{21} & A^{22} \end{bmatrix} 
= \frac{1}{2}\Big\{ \frac{1}{2}  \begin{bmatrix} 1 & 1\\1 & 1 \end{bmatrix} 
\otimes  \frac{1}{2}\begin{bmatrix} 1&i\\-i&1 \end{bmatrix} \Big\} +
\frac{1}{2}\Big\{  \frac{1}{2}  \begin{bmatrix} 1 & 1\\1 & 1 \end{bmatrix} 
\otimes \frac{1}{2}  \begin{bmatrix} 1 & 1\\1 & 1 \end{bmatrix}\Big\} ,$ 
where $A^{11}=  \begin{bmatrix}	2 & 1+i  \\ 1-i & 2 \end{bmatrix} $ and 
$A^{12}= \begin{bmatrix} 2 & 1+i  \\ 1-i & 2 \end{bmatrix}
=  \begin{bmatrix} 1&1\\1&1\end{bmatrix}  
+ \begin{bmatrix} 1 & i\\-i & 1 \end{bmatrix} $, \\
and $A^{xy} A^{mn}=A^{mn}A^{xy}$.
								
$$	
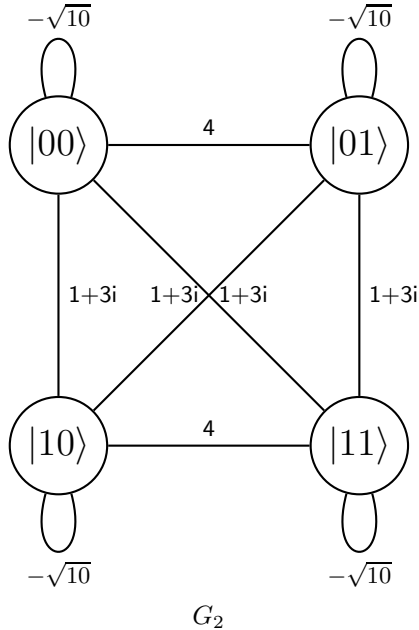
\begin{tikzpicture}[auto, node distance=4cm, every loop/.style={},
thick,main node/.style={circle,draw,font=\sffamily\Large\bfseries}]
\node[main node] (1) {$\ket{00}$};
\node[main node] (2) [ right of=1] {$\ket{01}$};
\node[main node] (3) [ below of=1] {$\ket{10}$};
\node[main node] (4) [ below of=2] {$\ket{11}$};
\draw (2, -6) node[below] { $ \hspace{1cm} G_2 \hspace{1cm}$};   	    	 
\path[every node/.style={font=\sffamily\small}]
(1) edge node  [] {4} (2)
(1) edge node  [] {1+3i} (3)
(1) edge node  [left] {1+3i} (4)
(1) edge [loop above] node[]{ $-\sqrt{10}$} (1)
(2) edge node  [right] {1+3i} (3)
(2) edge node  [] {1+3i} (4)	
(2) edge [loop above] node[]{$-\sqrt{10}$} (2)
(3) edge node  [] {4} (4) 
(3) edge [loop below] node[]{ $-\sqrt{10}$} (3)
(4) edge [loop below] node[]{ $-\sqrt{10}$} (4);
\end{tikzpicture} $$
									
$\rho_{G_2 }= \frac{1}{16} \begin{bmatrix} 4 & 4 & 1+3i & 1+3i \\ 4 & 4 & 1+3i & 1+3i \\
1-3i & 1-3i & 4 & 4 \\ 1-3i  & 1-3i & 4 & 4\end{bmatrix} 
= \begin{bmatrix} A^{11} & A^{12}\\ A^{21} & A^{22} \end{bmatrix} 
= \frac{12}{16}\Big\{ \frac{1}{2}\begin{bmatrix} 1&i\\-i&1 \end{bmatrix} 
\otimes \frac{1}{6}  \begin{bmatrix} 3 & 3\\3 & 3 \end{bmatrix} \Big\} 
+ \frac{4}{16}\Big\{ \frac{1}{2}  \begin{bmatrix} 1 & 1\\1 & 1\end{bmatrix} 
\otimes \frac{1}{2}  \begin{bmatrix} 1 & 1\\1 & 1 \end{bmatrix}\Big\} ,$ 
where $A^{11}=  \begin{bmatrix} 4 & 4 \\4 & 4 \end{bmatrix} $ 
and $A^{12}= \begin{bmatrix} 1+3i & 1+3i \\1+3i  & 1+3i \end{bmatrix}
=  \begin{bmatrix} 1&1\\1&1\end{bmatrix}  + i\begin{bmatrix}3& 3\\ 3&3\end{bmatrix} $, \\
and  $A^{xy} A^{mn}=A^{mn}A^{xy}$.

\begin{theorem}
Let $G$ be a weighted graph on $2^n$ vertices associated with an $n$-qubit quantum state and $E(G)$ be the edge set with $E(G)={(v_{ij}, v_{kl}) \mid \text{ $\exists$  as an edge between vertices } v_{ij} \text{ and } v_{kl}}$. For 
 $(v_{ij},v_{kl})\in E(G)$, if both $j$ and $l$ are odd or both $j$ and $l$ are even then the graph $G$ is associated with a separable state.
\end{theorem} 
Proof  Let us consider a weighted graph $G$ with $2^n$ vertices, where each vertex is associated with an $n$-qubit quantum state. 							
If $(v_{ij},v_{kl})\in E(G)$ such that either both $j$ and $l$ are odd or both $j$ and $l$ are even, then all blocks of the density operator ($\rho_G=  [A^{xy}]_{2^q 2^q}$, and $(A^{xy}= [a^{xy}_{ij}]_{2^p \times 2^p} )$) associated with the graph $G$ are diagonal, which clearly shows that blocks commute.
Using Theorem (\ref{TH3}), the  graph $G$ is associated with a separable state if blocks commute. Hence proved.

\textbf{Example:} The following graph $G$ further illustrates an example of the theorem.

$$	\begin{tikzpicture}[auto, node distance=2.5cm, every loop/.style={},
thick,main node/.style={circle,draw,font=\sffamily\Large\bfseries}]
\node[main node] (1) {$v_{11}$};	
\node[main node] (2) [ right of=1] {$v_{12}$};
\node[main node] (3) [ right of=2] {$v_{13}$};
\node[main node] (4) [ right of=3] {$v_{14}$};
\node[main node] (5) [ below of=1] {$v_{21}$};
\node[main node] (6) [ below of=2] {$v_{22}$};
\node[main node] (7) [ below of=3] {$v_{23}$};
\node[main node] (8) [ below of=4] {$v_{24}$};
\draw (4,-4) node[below] { $ \hspace{1cm} G \hspace{1cm}$};
\path[every node/.style={font=\sffamily\small}]
(1) edge [] node[] {} (7)
(2) edge [] node[] {} (8)
(3) edge [] node[] {} (5)
(3) edge [] node[] {} (7);
\end{tikzpicture} $$ 

 $\rho_G = \frac{1}{7}\begin{bmatrix} 1 & 0 & 0 & 0 & 0 & 0 & -1 & 0 \\ 
 0 & 1 & 0 & 0 & 0 & 0 & 0 & -1 \\  0 & 0 & 2 & 0 & -1 & 0 & -1 & 0\\
 0 & 0 & 0 & 0 & 0 & 0 & 0 & 0 \\ 0 & 0 & -1 & 0 & 1 & 0 & 0 & 0 \\ 
 0 & 0 & 0 & 0 & 0 & 0 & 0 & 0 \\-1 & 0 & -1 & 0 & 0 & 0 & 2 & 0 \\ 
 0 & 0 & 0 & 0 & 0 & 0 & 0 & 0 \end{bmatrix} $

One can see that all two-dimensional blocks within the density operator are diagonal, and therefore exhibit commutativity. For weighted graph also all blocks will be diagonal and hence commute.

\begin{theorem}
Let $G$ be a weighted graph on $2^n$ vertices associated with an $n$-qubit quantum state and $E(G)$ be the edge set with $E(G)={(v_{ij}, v_{kl}) \mid \text{ $\exists$  as an edge between vertices } v_{ij} \text{ and } v_{kl}}$. If $(v_{ij},v_{kl})\in E(G)$ for all $i=k$ then the graph $G$ is associated with  a  separable state.
\end{theorem} 
Proof  Let $G$ be a weighted graph on $2^n$ vertices that corresponds to an $n$-qubit quantum state. The edge set of $G$, denoted by $E(G)$, consists of all pairs of vertices $(v_{ij},v_{kl})$ that are connected by an edge. If each pair of vertices with $i=k$ is connected by an edge then the density operator associated with the graph $G$ can be expressed by a diagonal block matrix. This further shows that $ (\overline{\text{Partial}}U) \rho_G = \rho_G $ for $U= \underbrace{ I_2 \otimes \dots \otimes I_2}_{q} \otimes {U_{q+1} \otimes \dots \otimes U_{q+p} }$, where $U_{q+1} , \dots, U_{q+p}  $ are unitary operators (either $I_2$ or $\sigma_x$), and $A^{xy} \geq0 $ then blocks commute.  Using Theorem (\ref{TH3}), the  graph $G$ is associated with a separable state.
Hence the proof.

\textbf{Example:}  The graph $G$ below serves as an example of the theorem.
$$	\begin{tikzpicture}[auto, node distance=2.2cm, every loop/.style={},
    thick,main node/.style={circle,draw,font=\sffamily\Large\bfseries}]
   \node[main node] (1) {$v_{11}$};
   \node[main node] (2) [ right of=1] {$v_{12}$};
   \node[main node] (3) [ right of=2] {$v_{21}$};
   \node[main node] (4) [ right of=3] {$v_{22}$};
   \node[main node] (5) [ below of=1] {$v_{13}$};
   \node[main node] (6) [ below of=2] {$v_{14}$};
   \node[main node] (7) [ below of=3] {$v_{23}$};
   \node[main node] (8) [ below of=4] {$v_{24}$};
  \draw (3.5,-4) node[below] { $ \hspace{1cm} $G$ \hspace{1cm}$};
  \path[every node/.style={font=\sffamily\small}]
  (1) edge [] node[] {} (2)
  edge [left] node[] {} (6)
  edge [] node[] {} (5)
  (2) edge [right] node[] {} (5)
      edge [] node[] {} (6)
  (3) edge [] node[] {} (4)
      edge  node[] {} (7)
      edge  node[ left] {} (8)
  (4) edge [right] node[] {} (7)
      edge  node[] {} (8)
 (5) edge [] node[] {} (6)
 (7) edge  node[] { } (8);
\end{tikzpicture}$$	    

 $\rho_G = \frac{1}{24} \begin{bmatrix} 3 & -1 & -1 & -1 & 0 & 0 & 0 & 0 \\ 
  -1 & 3 & -1 & -1 & 0 & 0 & 0 & 0  \\  -1 & -1 & 3 & -1 & 0 & 0 & 0 & 0 \\
  -1 & -1 & -1 & 3 & 0 & 0 & 0 & 0  \\ 0 & 0 & 0 & 0 & 3 & -1 & -1 & -1 \\ 
 0 & 0 & 0 & 0 & -1 & 3 & -1 & -1  \\ 0 & 0 & 0 & 0 & -1 & -1 & 3 & -1  \\ 
 0 & 0 & 0 & 0 & -1 & -1 & -1 & 3 \end{bmatrix} $
Here, the density operator is a diagonal block matrix featuring positive semi-definite diagonal blocks and zero matrices for non-diagonal blocks. On applying weight to this type of graphs, positive semi-definite diagonal blocks persist and non-diagonal blocks remain zero matrices.
												
\begin{corollary}
Let $G $ be a simple graph with $2^n$ vertices associated with an $n$-qubit state. For a unitary operator  $U= I_2 \otimes {U_2 \otimes \dots \otimes U_n}$ where $U_2, \dots, U_n $ are unitary operator (either $I_2$ or $\sigma_x$), if $G=(PartiallyU)G$ then the graph is associated with a separable state.
\end{corollary}
Proof. Let $G $ be a simple graph with $2^n$ vertices associated with $n$-qubit state. If $G = (\text{Partial}U)G$ then $ (Ub_i,Ub_j) \in E(G)$ for all $(b_i,b_j) \in E(G)$, where
$U_k= \begin{cases}
I_2  \text{ if } {c_k}^i= {c_k}^j\\
\sigma_x \text{ if } {c_k}^i \neq {c_k}^j
\end{cases}$ 
which shows that blocks are symmetric and can be expressed as 
$A^{xy} = B^{xy}+C^{xy}$ for $x \neq y$ where$ B^{xy}$ is positive semi-definite  and $C^{xy}$ is negative semi-definite. Clearly, $A_{xx}- \sum_{y \neq x}B_{xy}- \sum_{y \neq x}C_{xy} \geq 0$ for all $x$, therefore the graph $G$ is associated with a separable state from corollary \ref{C1} \cite{braunstein2006some,dutta2016bipartite,joshi2022entanglement,joshi2018concurrence,wu2006conditions}.
													
\textbf{Example: } The graph $G$ below is an example that illustrates the previously mentioned corollary.
$$	\begin{tikzpicture}[auto, node distance=2.5cm, every loop/.style={},
thick,main node/.style={circle,draw,font=\sffamily\Large\bfseries}]
\node[main node] (1) {$\ket{000}$};	\node[main node] (2) [ right of=1] {$\ket{001}$};
\node[main node] (3) [ right of=2] {$\ket{010}$};
\node[main node] (4) [ right of=3] {$\ket{011}$};
\node[main node] (5) [ below of=1] {$\ket{100}$};
\node[main node] (6) [ below of=2] {$\ket{101}$};
\node[main node] (7) [ below of=3] {$\ket{110}$};
\node[main node] (8) [ below of=4] {$\ket{111}$};
\draw (4,-4) node[below] { $ \hspace{1cm} G \hspace{1cm}$};
\path[every node/.style={font=\sffamily\small}]
(1) edge [] node[] {} (5)
(1) edge [] node[] {} (8)
(2) edge [] node[] {} (6)
(3) edge [] node[] {} (7)
(4) edge [] node[] {} (5)
(4) edge [] node[] {} (8);
\end{tikzpicture} $$ 
$$\downarrow_{partial(I_2 \otimes U_2 \otimes U_3})$$
$$ \begin{tikzpicture}[auto, node distance=2.5cm, every loop/.style={},
thick,main node/.style={circle,draw,font=\sffamily\Large\bfseries}]
\node[main node] (1) {$\ket{000}$};
\node[main node] (2) [ right of=1] {$\ket{001}$};
\node[main node] (3) [ right of=2] {$\ket{010}$};
\node[main node] (4) [ right of=3] {$\ket{011}$};
\node[main node] (5) [ below of=1] {$\ket{100}$};
\node[main node] (6) [ below of=2] {$\ket{101}$};
\node[main node] (7) [ below of=3] {$\ket{110}$};
\node[main node] (8) [ below of=4] {$\ket{111}$};
\draw (4,-4) node[below] { $ \hspace{1cm} G \hspace{1cm}$};
\path[every node/.style={font=\sffamily\small}]
(1) edge [] node[] {} (5)
(1) edge [] node[] {} (8)
(2) edge [] node[] {} (6)
(3) edge [] node[] {} (7)
(4) edge [] node[] {} (5)
(4) edge [] node[] {} (8);
\end{tikzpicture} $$																
\begin{corollary}
Let the density operator $\rho_G$ be a block matrix with dimension $2^q \times 2^q $ of a simple graph $G$ on $2^{n=p+q}$ vertices, and $V=\{ b_i={c_1}^i\otimes {c_2}^i \otimes \dots \otimes {c_n}^i \text{ for all $i= 1,2,\dots ,2^n$}\}$ be the vertex set where $c_i's$ are  column vectors in ${\mathbb{C}}^2$. Considering the unitary operator $U= \underbrace{ I_2 \otimes \dots \otimes I_2}_{q} \otimes {U_{q+1} \otimes \dots \otimes U_{q+p} }$, where $U_{q+1} , \dots, U_{q+p}  $ are unitary operators (either $I_2$ or $\sigma_x$), if $(\text{Partial}U) \rho_G=\rho_G$ then the graph is associated with a separable state in ${\mathbb{C}}^{2^p} \otimes {\mathbb{C}}^{2^q}$.
\end{corollary}															
Proof Let  $\rho_G= \frac{1}{L(G)} [l_{ij} b_i \otimes {b_j}^T]$ be a density operator of a simple graph $G$ on $2^{n=p+q}$ vertices, and $V=\{ b_i={c_1}^i\otimes {c_2}^i \otimes \dots \otimes {c_n}^i \text{ for all $i= 1,2,\dots ,2^n$}\}$ be the vertex set. The partial quantum gate on the density is define as $(\text{Partial}U) \rho_G= \frac{1}{L(G)} [l_{ij} Ub_i \otimes {Ub_j}^T]$, where 
$U_k= \begin{cases}
I_2  \text{ if } {c_k}^i= {c_k}^j\\
\sigma_x \text{ if } {c_k}^i \neq {c_k}^j
\end{cases}$. 
Therefore, $(\text{Partial}U) \rho_G = \rho_G $ clearly shows that all  blocks of $ \rho_G $ are symmetric implying that the density operator can be expressed in the product form by corollary \ref{C1}. Hence the proof \cite{braunstein2006some,dutta2016bipartite,joshi2022entanglement,joshi2018concurrence,wu2006conditions}.
																
\begin{corollary}
Let $\rho_G $ be a density operator of a simple graph with $2^n$ vertices associated with an $n$-qubit state. For a unitary operator $ U= I_2 \otimes \dots \otimes I_2 \otimes \sigma_x $, if $\rho_G =(PartialU) \rho_G$ then the graph is associated with a separable state in ${\mathbb{C}}^2 \otimes {\mathbb{C}}^{2^{(n-1)}}$.
\end{corollary} 
Proof The density operator can also be expressed as	$$\rho_G= \rho_{ij} b_i \otimes {b_j}^T,$$ where $b_i= {c_1}^i \otimes {c_2}^i \otimes \dots \otimes {c_n}^i$ , are basis  vectors in ${\mathbb{C}}^{2^n}$ and $c_i$ are basis vectors in ${\mathbb{C}}^{2}$. Here
$$ {\rho_G}'=(PartialU) \rho_G = \begin{cases}
\rho_{ij} b_i \otimes {b_j}^T ~~~~ {c_n}^i = {c_n}^j \\\rho_{ij}{ (Ub_i) } \otimes {(Ub_j)}^T ~~~~ {c_n}^i \neq {c_n}^j
\end{cases}.$$
If $\rho_G= {\rho_G}'$ then $Tr(\rho_G) = Tr({\rho_G}')$ =1 and $\sum_{i=1} \rho_{ij}= \sum_{i=1} {\rho_{ij}}' =0$. Further, $\rho_G$ is a block matrix with dimension $2^{(n-1)} \times 2^{(n-1)} $ suggesting that each block is symmetric. Therefore, the density operator of a simple graph $G$ is separable in ${\mathbb{C}}^2 \otimes {\mathbb{C}}^{2^{(n-1)}}$ by corollary \ref{C1} \cite{braunstein2006some,dutta2016bipartite,joshi2022entanglement,joshi2018concurrence,wu2006conditions}.
																	
\begin{corollary}
Let $G$ be a simple graph on $2^n$ vertices associated with an $n$-qubit state. Assume $V(G)=\{v_{ij} : i=1,2,\dots,2^p \text{and} j=1,2,\dots,2^q\}$ be a vertex set and $E(G)$ be an edge set. If $(v_{ij},v_{kl}) \in E(G)$ such that there exists $(v_{i(j+1)},v_{k(l-1)}) \in E(G)$ where $j$ is odd and $l$ is even then the graph $G$ is associated with a separable state. 
\end{corollary}
Proof 	For $(v_{ij},v_{kl}) \in E(G)$, there exists $(v_{i(j+1)},v_{k(l-1)}) \in E(G)$ such that $j$ is odd and $l$ is even, we have $ {(\text{Partial}U) \rho_G} = \rho_G $ with $U= \underbrace{ I_2 \otimes \dots \otimes I_2}_{n-1} \otimes {U_{n}} $ where  $U_{n}$ is a unitary operator (either $I_2$ or $\sigma_x$). Therefore, the blocks of density operator are symmetric and can be decomposed as $A_{ij}= B_{ij}- C_{ij}$ where $B_{ij}$ and $C_{ij}$ are positive semi definite matrices, and $A_{ii}- \sum_{j \neq i}B_{ij}- \sum_{j \neq i}C_{ij} \geq 0$ for all $i$. Hence proved \cite{braunstein2006some,dutta2016bipartite,joshi2022entanglement,joshi2018concurrence,wu2006conditions}.
																	
\begin{theorem}
Let $G$ be a weighted graph on $2^n$ vertices associated with $(n=p+q)$-qubit  state and $V(G)=\{v_{ij} \text{ for all } i=1,2,\dots,2^p \text{ and } j=1,2,\dots,2^q\}$ be a vertex set with $N(v_{ij}) = \{v_{kl} \text{ such that } (v_{ij},v_{kl}) \in E(G)\}$ be a neighbouring set of vertex $v_{ij}$. If $v_{ij} \in N(v_{kl})$ for $i=k$ then the graph $G$ is associated with a separable state.
\end{theorem}
Proof. 	Let $G$ be a weighted graph on $2^n$ vertices associated with a $(n=p+q)$-qubit  state. We define $N(v_{ij}) = \{v_{kl} \text{ such that } (v_{ij,v_{kl}}) \in E(G)\}$ to be a neighbouring set of vertex $v_{ij}$ of $G$ associated with an $(n)$-qubit mixed state. If $v_{ij} \in N(v_{kl})$ for $i=k$ then the density operator is a diagonal block matrix. Using theorem \ref{TH3}, $\rho_G$ can be written as a product state. Hence the proof.

Example. Consider the following graph $G$ on $8$ vertices associated with a separable state.
$$	
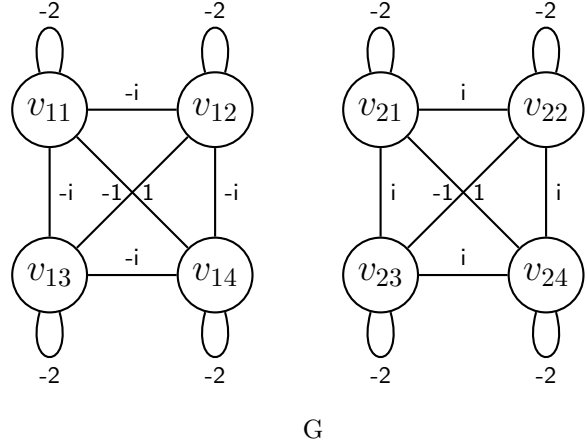
\begin{tikzpicture}[auto, node distance=2.2cm, every loop/.style={},
    thick,main node/.style={circle,draw,font=\sffamily\Large\bfseries}]
  \node[main node] (1) {$v_{11}$};
   \node[main node] (2) [ right of=1] {$v_{12}$};
   \node[main node] (3) [ right of=2] {$v_{21}$};
   \node[main node] (4) [ right of=3] {$v_{22}$};
   \node[main node] (5) [ below of=1] {$v_{13}$};
   \node[main node] (6) [ below of=2] {$v_{14}$};
   \node[main node] (7) [ below of=3] {$v_{23}$};
   \node[main node] (8) [ below of=4] {$v_{24}$};
  \draw (3.5,-4) node[below] { $ \hspace{1cm} $G$ \hspace{1cm}$};
  \path[every node/.style={font=\sffamily\small}]
  (1) edge [] node[] {-i} (2)
  edge [left] node[] {-1} (6)
  edge [] node[] {-i} (5)
  edge [loop above] node[] {-2} (1)
  (2) edge [right] node[] {1} (5)
      edge [] node[] {-i} (6)
      edge [loop above ] node[] {-2} (2)
  (3) edge [] node[] {i} (4)
      edge  node[] { i} (7)
     edge  node[ left] { -1} (8)
      edge  [loop above ]node { -2} (3)
  (4) edge [right] node[] {1} (7)
      edge  node[] { i} (8)
      edge  [loop above ]node { -2} (4)
  (5) edge [] node[] {-i} (6)
      edge [loop below] node[] {-2} (5)
  (6) edge [loop below] node[] {-2} (6)
  (7) edge  node[] { i} (8)
     edge  [loop below ]node { -2} (7)
  (8) edge  [loop below ]node { -2} (8);
\end{tikzpicture}$$																

\begin{theorem}
Let $G$ be a simple graph on $2^n$ vertices associated with a $(n=p+q)$-qubit state. Let $V(G)=\{v_{ij} : i=1,2,\dots,2^p \text{and} j=1,2,\dots,2^q\}$ be a vertex set and $N(v_{ij}) = \{v_{kl} \text{ ; } (v_{ij},v_{kl}) \in E(G)\}$ be a neighbouring set to vertex $v_{ij}$. If $v_{ij} \in N(v_{kl}) $ then there exists $v_{il} \in N(v_{kj}) $ for all $i$,$k$ such that the graph $G$ is associated with a separable state.
\end{theorem}
Proof 	Let $G$ be a simple graph on $2^n$ vertices associated with a $(n=p+q)$-qubit state.   If $v_{ij} \in N(v_{kl}) $ and there exists $v_{il} \in N(v_{kj}) $ for all $i$,$k$ then the blocks of density operator are symmetric. Therefore, the graph $G$ is associated wih a separable state by corollary \ref{C1} \cite{braunstein2006some,dutta2016bipartite,joshi2022entanglement,joshi2018concurrence,wu2006conditions}. 

Example:  Consider a graph $G$ on $8$ vertices associated with a separable state.
$$	
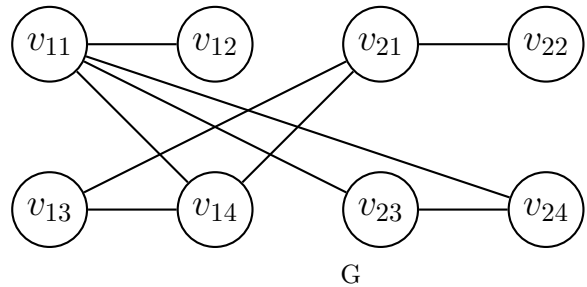
\begin{tikzpicture}[auto, node distance=2.2cm, every loop/.style={},
thick,main node/.style={circle,draw,font=\sffamily\Large\bfseries}]
\node[main node] (1) {$v_{11}$};
   \node[main node] (2) [ right of=1] {$v_{12}$};
   \node[main node] (3) [ right of=2] {$v_{21}$};
   \node[main node] (4) [ right of=3] {$v_{22}$};
   \node[main node] (5) [ below of=1] {$v_{13}$};
   \node[main node] (6) [ below of=2] {$v_{14}$};
   \node[main node] (7) [ below of=3] {$v_{23}$};
   \node[main node] (8) [ below of=4] {$v_{24}$};
\draw (4,-2.8) node[below] { $ \hspace{1cm} $G$ \hspace{1cm}$};
\path[every node/.style={font=\sffamily\small}]
(1) edge [] node[] {} (2)
 edge [] node[] {} (6)
 edge [] node[] {} (8)
 edge [] node[] {} (7)
(3) edge [] node[] {} (4)
(3)edge [] node[] {} (5)
(3)edge [] node[] {} (6)
(5) edge [] node[] {} (6)
(7) edge  node[] { } (8);
\end{tikzpicture}$$
Clearly, $N(11)= \{12,14,23,24\}$, $N(12)= \{11\}$,$N(13)= \{14,21\}$,$N(14)= \{11,13,21\}$, $N(21)= \{13,14,22\}$, $N(22)= \{21\}$,$N(23)= \{11,24\}$,$N(24)= \{11,23\}$.

\begin{theorem}
Let $G$ be a simple graph on $2^n$ vertices associated with a $(n=p+q)$-qubit mixed state and $V(G)=\{v_{ij} \text{ for all } i=1,2,\dots,2^p \text{ and } j=1,2,\dots,2^q\}$ be a vertex set such that $N(v_{ij}) = 
\{v_{kl} \text{ ; } (v_{ij,v_{kl}}) \in E(G)\}$ is a neighbouring set to vertex $v_{ij}$. If $ v_{ij} \in N(v_{kl}) $ for all $i \neq k$, there exists either $ v_{i(j-1)} \in N(v_{k(l+1)}) $ for all even $j$ and odd $l$, or  $ v_{i(j+1)} \in N(v_{k(l-1)}) $ for all odd $j$ and even $l$ then the graph $G$ is associated with a separable state.
\end{theorem}

Proof 
Let $G$ be a simple graph on $2^n$ vertices associated with a $(n=p+q)$-qubit mixed state. If $ v_{ij} \in N(v_{kl}) $ for all $i \neq k $ then there exists either $ v_{i(j-1)} \in N(v_{k(l+1)}) $ for all even $j$ and odd $l$ or   $ v_{i(j+1)} \in N(v_{k(l-1)}) $ for all even $l$ and odd $j$ showing that each $2 \times 2$ off diagonal blocks of order $2^p \times 2^p$  are symmetric and can be expressed as $$\rho_{G}= \frac{1}{Tr(L(G))} \begin{bmatrix}
	B_{11} & B_{12} & . & . & .	&B_{12^p}\\
	. & . & . & . & .	& . \\
	. & . & . & . & .	& . \\
	. & . & . & . & .	& .\\
	B_{2^p1} & B_{2^p2} & . & . & .	& B_{2^p2^p}
\end{bmatrix}$$
Here $B_{ij}$ are block matrices and $B_{ij} = C_{ij}+ D_{ij}$ are written as sum of positive semi-definite and negative semi-definite matrices implying $ B_{ii}- \sum_{j} C_{ij} -\sum_{j}D_{ij} \geq 0$ which shows that $\rho_G$ can be written as a product state. Hence proved \cite{braunstein2006some,dutta2016bipartite,joshi2022entanglement,joshi2018concurrence,wu2006conditions}.

Example. Consider the following graph $G$ on $8$ vertices associated with a separable state.
$$	
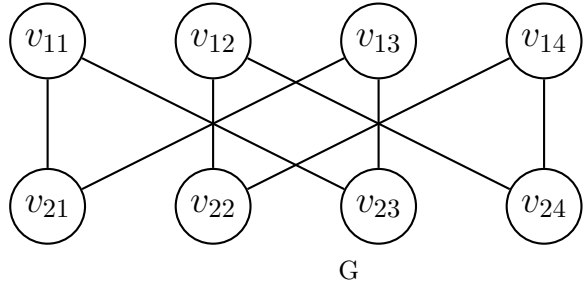
\begin{tikzpicture}[auto, node distance=2.2cm, every loop/.style={},
thick,main node/.style={circle,draw,font=\sffamily\Large\bfseries}]
\node[main node] (1) {$v_{11}$};
   \node[main node] (2) [ right of=1] {$v_{12}$};
   \node[main node] (3) [ right of=2] {$v_{13}$};
   \node[main node] (4) [ right of=3] {$v_{14}$};
   \node[main node] (5) [ below of=1] {$v_{21}$};
   \node[main node] (6) [ below of=2] {$v_{22}$};
   \node[main node] (7) [ below of=3] {$v_{23}$};
   \node[main node] (8) [ below of=4] {$v_{24}$};
\draw (4,-2.8) node[below] { $ \hspace{1cm} $G$ \hspace{1cm}$};
\path[every node/.style={font=\sffamily\small}]
(1) edge [] node[] {} (5)
edge [] node[] {} (7)
(2) edge [] node[] {} (6)
edge  node[] { } (8)
(3) edge [] node[] {} (5)
edge [] node[] {} (7)
(4) edge [] node[] {} (6)
 (4) edge  node[] { } (8);
\end{tikzpicture}$$

\begin{theorem}
Let $G$ be a simple graph on $2^n$ vertices associated with a $(n=p+q)$-qubit mixed state and $V(G)=\{v_{ij} \text{ for all } i=1,2,\dots,2^p \text{ and } j=1,2,\dots,2^q\}$ be a vertex set such that $N(v_{ij}) = \{v_{kl} \text{ ; } (v_{ij,v_{kl}}) \in E(G)\}$ be the neighbouring set to vertex $v_{ij}$. If $ v_{ij} \in N(v_{kl}) $ for all $i, k$, then there exists $ v_{i(j-1)} \in N(v_{k(l+1)}) $ for all even $j$ and odd $l$ or  $ v_{i(j+1)} \in N(v_{k(l-1)}) $ for all even $l$ and odd $j$ and the graph $G$ is associated with a separable state .
\end{theorem}

Proof Let $G$ be a simple graph on $2^n$ vertices associated with a $(n=p+q)$-qubit mixed state.   If $ v_{ij} \in N(v_{kl}) $ for all $i, k $, then  there exists $ v_{i(j-1)} \in N(v_{k(l+1)}) $ for all even $j$ and odd $l$ or $ v_{i(j+1)} \in N(v_{k(l-1)}) $ for all even $l$ and odd $j$ suggesting that each $2 \times 2$ blocks of $\rho_G$ are symmetric. Hence the proof \cite{joshi2022entanglement,joshi2018concurrence,wu2006conditions}.

\begin{theorem} \label{T8}
Let $G$ be simple a graph on $2^n$ vertices associated with a $(n=p+q)$-qubit state and $V(G)=\{v_{ij} \text{ for all } i=1,2,\dots,2^p \text{ and } j=1,2,\dots,2^q\}$ be a vertex set such that  $N(v_{ij}) = \{v_{kl} \text{ ; } (v_{ij,v_{kl}}) \in E(G)\}$ be a neighbouring set to vertex $v_{ij}$. Assume $N(G) =\{N(v_{ij})  \text{ for all } i,j \}  \}$. If there exists an equivalence relation $R$ on the set $N(G)$ then the graph is associated with a separable state where the relation R on the neighbourhood set of vertex is defined as ${N(v_{ij})} R {N(v_{kl})}  \longleftrightarrow \card{N(v_{ij})} = \card{N(v_{kl})}$.
\end{theorem}

Proof 	Let $G$ be a simple graph on $2^n$ vertices associated with a $(n=p+q)$-qubit state. Clearly, a relation $R$ on  the set $N(G) =\{N(v_{ij})  \text{ for all } i,j \}  \}$ is equivalence if it is reflexive, i.e. $N(v_{ii}) R N(v_{ii})$ for all $i$ or symmetric, i.e.  if $N(v_{ij}) R N(v_{kl})$  then  $N(v_{kl}) R N(v_{ij})$, or transitive, i.e. if $N(v_{ij}) R N(v_{kl})$  and   $N(v_{kl}) R N(v_{rs})$ then  $N(v_{ij}) R N(v_{rs})$. This shows that $G$ is a complete graph. Hence proved \cite{braunstein2006laplacian,hassan2007combinatorial,joshi2022entanglement,joshi2018concurrence,wu2006conditions}. 

Example: Consider the following graph $G$ on $4$ vertices associated with a separable state.
$$	\begin{tikzpicture}[auto, node distance=4cm, every loop/.style={},
thick,main node/.style={circle,draw,font=\sffamily\Large\bfseries}]
\node[main node] (1) {$v_{11}$};
   \node[main node] (2) [ right of=1] {$v_{12}$};
   \node[main node] (3) [ below of=1] {$v_{21}$};
   \node[main node] (4) [ below of=2] {$v_{22}$};
\draw (2, -5) node[below] { $ \hspace{1cm} G \hspace{1cm}$};   	    	 
\path[every node/.style={font=\sffamily\small}]
(1) edge node[] {} (2)
 edge node  [] {} (3)
  edge node  [] {} (4)
(2) edge node[] {} (4)
  edge node  [] {} (3)
(3) edge node[] {} (4);
\end{tikzpicture}$$

\begin{corollary}
Every complete graph and regular graph is separable under equivalence relation of neighbouring set, where the relation R on vertex set is defined as ${N(v_{ij})} R {N(v_{kl})}  \longleftrightarrow \card{N(v_{ij})} = \card{N(v_{kl})}$.
\end{corollary}

\begin{theorem} \label{T9}
Let $G$ be a simple graph on $2^n$ vertices and $N(v_{ij})= \{v_{kl} \text{ such that } (v_{ij},v_{kl}) \in E(G) \} $. Assuming $\card{N(v_{ij})}=n$ and $\card{N(v_{kl})}=m$, the graph $G$ is separable if there exists an equivalence relation $ R$ on neighbourhood vertex set, defined as ${N(v_{ij})} R {N(v_{kl})}  \longleftrightarrow  \begin{cases} \card{N(v_{ij})} = n \text{ and } \card{N(v_{kl})}=m , \text{ for } i\neq k \\ \card{N(v_{ij})} =  \card{N(v_{kl})}, \text{ otherwise } \end{cases}$ 
\end{theorem}
Proof: Theorem (\ref{T9}) shares a common method of proof with theorem (\ref{T8}), even though theorem (\ref{T8}) and theorem (\ref{T9}) address entirely different propositions.

\begin{theorem}
Let $G$ be a weighted graph on $2^n$ vertices and $\rho_G=[A^{xy}]_{2^q \times 2^q}$ be a density operator. Suppose a relation R on a set $ A=\{ A^{xy} \} $ such that $R$ is defined as $A^{ij} R A^{kl}$ if and only if $A^{ij} A^{kl} = A^{kl} A^{ij} $. Graph $G$ will be associated with a  separable state if $R$ is an equivalence relation.
\end{theorem}
Proof  Let $G$ be a weighted graph on $2^n$ vertices and $\rho_G=[A^{xy}]_{2^q \times 2^q}$ be a density operator. Considering R on a set $ A=\{ A^{xy} \} $ is defined as $A^{ij} R A^{kl}$ if and only if $A^{ij} A^{kl} = A^{kl} A^{ij} $, $R$ is reflexive, symmetric and transitive which shows that each block of the density operator commute with each other. By theorem \ref{TH3}, the graph $G$ is associated with a separable state.

\begin{theorem}
	 Let $G$ be a simple graph on $2^n$ vertices and $N(v_{ij})= \{v_{kl} \text{ such that } (v_{ij},v_{kl}) \in E(G) \} $. Assume that the neighbourhood sequence of vertices of a graph $G$
	is $ N_1 \leq N_2 \leq \dots \leq  N_{2^n}$ and eigen value sequence  of density operator $\rho_G$ is  $ \lambda_1 \leq \lambda_2 \leq \dots \leq  \lambda_{2^n}$. 
	If $ \lambda_{2^n} = \frac{N_1}{\card{E(G)}}$ then the graph $G$ is associated with a separable $n$-qubit state.
\end{theorem}
Proof  Let $G$ be a simple graph on $2^n$ vertices and $N(v_{ij})= \{v_{kl} \text{ such that } (v_{ij},v_{kl}) \in E(G) \} $. The neighbourhood sequence of vertices  of $G$
is $ N_1 \leq N_2 \leq \dots \leq  N_{2^n}$ and eigen value sequence  of density operator $\rho_G$ is  $ \lambda_1 \leq \lambda_2 \leq \dots \leq  \lambda_{2^n}$. We know that 
$\lambda_{2^n} \leq \frac{1}{2\card{E(G)}} [N_1 + \frac{1}{2} + \sqrt{(N_1 - \frac{1}{2} )^2 +\sum_{i}^{2^n}N_i(N_i-N_{2^n})}]$ \cite{zhang2011laplacian}. If $ \lambda_{2^n} = \frac{N_1}{\card{E(G)}}$ then 
 $ N_1 = N_2 = \dots = N_{2^n}$ which clearly shows that the density operator can also be expressed as $$ \rho_G= \frac{1}{Tr(L(G))} \biggl\{\left[ \begin{array}{c|c}
A & 0\\
\hline
0&A
\end{array}  \right] + \left[ \begin{array}{c|c}
\alpha I & C\\
\hline
C &\alpha I
\end{array}  \right]\biggr\}. $$ Hence the proof \cite{dutta2016bipartite,joshi2022entanglement,joshi2018concurrence,wu2006conditions}.
\section{Conclusion}
The present article addresses the entanglement versus separability problem represented by graph Laplacian. For this, we introduced a necessary condition for separability of $n$-qubit quantum states within the composite Hilbert space $H=H_1 \otimes H_2 \otimes \dots \otimes H_n$, utilizing a combination of unitary operators and the neighborhood set of a graph with $2^n$ vertices. In the context of simple graphs, if each vertex's neighborhood has an equal number of vertices, the state is determined to be separable. Moreover, for a connected simple graph where the neighborhood sets of vertices feature varying numbers of vertices and an equivalence relation exists among these sets, the state is further considered as separable. Based on this discussion, it is evident that the presence or absence of edges in the graph plays a pivotal role in determining whether the quantum states are separable or entangled. Additionally, we explored separability in the context of equivalence relations.

\section{Appendix}
\subsection{In this section, we briefly discuss some standard examples of separable and entangled states.}
\begin{enumerate}
    \item Separable State:
\begin{enumerate}
    \item $\ket{\psi} = \ket{00}$\\
    The density operator is $\rho= \begin{bmatrix}
      1 & 0 & 0 & 0 \\  0 & 0 & 0 & 0 \\  0 & 0 & 0 & 0 \\  0 & 0 & 0 & 0 
    \end{bmatrix} = \begin{bmatrix} A^{11} & A^{12} \\ A^{21} & A^{22} \end{bmatrix}$\\
    Here we can  see that $A^{ii}- A^{ij} \geq 0$, therefore by theorem \ref{TH1} the state is separable.\\
    Moreover, blocks commute with each other, therefore, by theorem \ref{TH3} the state is separable.
    
     \item $\ket{\psi} = \ket{000}$\\
    The density operator is $\rho= \begin{bmatrix}
      1 & 0 & 0 & 0 & 0 & 0 & 0 & 0 \\  0 & 0 & 0 & 0 & 0 & 0 & 0 & 0 \\  0 & 0 & 0 & 0 & 0 & 0 & 0 & 0 \\  0 & 0 & 0 & 0 & 0 & 0 & 0 & 0 \\ 0 & 0 & 0 & 0 & 0 & 0 & 0 & 0 \\ 0 & 0 & 0 & 0 & 0 & 0 & 0 & 0 \\ 0 & 0 & 0 & 0 & 0 & 0 & 0 & 0 \\ 0 & 0 & 0 & 0 & 0 & 0 & 0 & 0 
    \end{bmatrix} = \begin{bmatrix} A^{11} & A^{12} \\ A^{21} & A^{22} \end{bmatrix}$\\
    Here one can  observe that $A^{ii}- A^{ij} \geq 0$, therefore by theorem \ref{TH1}, the state is separable.\\
    Also, blocks commute with each other, and therefore by theorem \ref{TH3}, the state is separable.
    
    \item $ \ket{\psi} =  cos \theta \ket{00} + sin \theta \ket{01}$\\
 The density operator is $\rho= \begin{bmatrix}
      cos^2\theta & cos\theta sin\theta & 0 & 0 \\  cos\theta sin\theta & sin^2 \theta & 0 & 0 \\  0 & 0 & 0 & 0 \\  0 & 0 & 0 & 0 
    \end{bmatrix} = \begin{bmatrix} A^{11} & A^{12} \\ A^{21} & A^{22} \end{bmatrix}$\\
    Here we can further visualize that $A^{ii}- A^{ij} \geq 0$, therefore, by theorem \ref{TH1}, given state is separable.\\
    Since blocks also commute with each other, hence by theorem \ref{TH3}, the state is separable.
    
    \item  $\ket{\psi} = \frac{1}{2} [\ket{00} + \ket{01} +\ket{10} +\ket{11}   ]$\\
     The density operator is $\rho= \frac{1}{4} \begin{bmatrix}
      1 & 1 & 1 & 1 \\  1 & 1 & 1 & 1 \\  1 & 1 & 1 & 1 \\  1 & 1 & 1 & 1 
    \end{bmatrix} = \begin{bmatrix} A^{11} & A^{12} \\ A^{21} & A^{22} \end{bmatrix}$
    Again, we have $A^{ii}- A^{ij} \geq 0$, therefore, by theorem \ref{TH1}, given state is separable.\\
    Since blocks also commute with each other, hence by theorem \ref{TH3}, the state is separable.
    
    \item $ \ket{\psi} =  cos \theta \ket{000} + e^{i\phi} sin \theta \ket{100}$
    
 The density operator can be represented as 
 $\rho= \begin{bmatrix}
 cos^2\theta& 0 & 0 & 0 & e^{i\phi}cos\theta sin\theta & 0 & 0 & 0 \\  0 & 0 & 0 & 0 & 0 & 0 & 0 & 0 \\  0 & 0 & 0 & 0 & 0 & 0 & 0 & 0 \\  0 & 0 & 0 & 0 & 0 & 0 & 0 & 0 \\ e^{-i\phi}cos\theta sin\theta & 0 & 0 & 0 & sin^2\theta & 0 & 0 & 0 \\ 0 & 0 & 0 & 0 & 0 & 0 & 0 & 0 \\ 0 & 0 & 0 & 0 & 0 & 0 & 0 & 0 \\ 0 & 0 & 0 & 0 & 0 & 0 & 0 & 0 
 \end{bmatrix} $
 $$=\begin{bmatrix} A^{11} & A^{12} \\ A^{21} & A^{22} \end{bmatrix} $$
 
    Clearly, $A^{ii}- A^{ij} \geq 0$, therefore by theorem \ref{TH1}, this state is separable.
    
      Considering that blocks also commute with each other, using theorem \ref{TH3}, the state is separable.
      
      \item $ \ket{\psi} =  cos \theta \ket{000} + e^{i\phi} sin \theta \ket{110}$\\
 The density operator in this case is $\rho= \begin{bmatrix}
 cos^2\theta& 0 & 0 & 0 & 0 & 0 & e^{i\phi}cos\theta sin\theta & 0 \\  0 & 0 & 0 & 0 & 0 & 0 & 0 & 0 \\  0 & 0 & 0 & 0 & 0 & 0 & 0 & 0 \\  0 & 0 & 0 & 0 & 0 & 0 & 0 & 0 \\ 0 & 0 & 0 & 0 & 0 & 0 & 0 & 0 \\ 0 & 0 & 0 & 0 & 0 & 0 & 0 & 0 \\ e^{-i\phi}cos\theta sin\theta & 0 & 0 & 0 & 0 & 0 & sin^2\theta & 0 \\ 0 & 0 & 0 & 0 & 0 & 0 & 0 & 0 
 \end{bmatrix} = \begin{bmatrix} A^{11} & A^{12} & A^{13}  &  A^{14} \\ A^{21} & A^{22} & A^{23}  &  A^{24} \\ A^{31} & A^{32} & A^{33}  &  A^{34} \\A^{41} & A^{42} & A^{43}  &  A^{44}  \end{bmatrix}$
    Since $A^{ii}- A^{ij} \geq 0$, therefore, by theorem \ref{TH1}, the state is separable.\\
    Also blocks commute with each other, therefore by theorem \ref{TH3}, the state is separable.
\end{enumerate}
    \item We now proceed to discuss a few entangled states based on our proposed criteria.
\begin{enumerate}
    \item $ \ket{\psi} = (1-a)\ket{00}\bra{00} + a \ket{\psi^+}\bra{\psi^+}$, where $\ket{\psi^+}= \frac{1}{\sqrt{2}}[\ket{01}+\ket{10}]$\\
     The density operator can be expressed as $\rho= \begin{bmatrix}
      1-a & 0 & 0 & 0 \\  0 & \frac{a}{2} & \frac{a}{2} & 0 \\  0 & \frac{a}{2} & \frac{a}{2} & 0 \\  0 & 0 & 0 & 0 
    \end{bmatrix} = \begin{bmatrix} A^{11} & A^{12} \\ A^{21} & A^{22} \end{bmatrix}$
    
   \item 
   $\rho_{MEMS} = \begin{bmatrix} 2 \delta & 0 & 0 &\frac{\delta}{2} \\ 
   0 & 1-2z(\delta) & 0 & 0 \\ 0 & 0 & 0 & 0 \\ \frac{\delta}{2} & 0 & 0 & 2 \delta  \end{bmatrix} =  \begin{bmatrix} A^{11} & A^{12} \\ A^{21} & A^{22} \end{bmatrix}$, 
   where $z( \delta )= \begin{cases} \frac{1}{3}, \delta <  \frac{2}{3}  \\ \frac{\delta}{2}, \delta \geq \frac{2}{3} \end{cases}$

\item $GHZ =cos \theta \ket{000} + e^{i\phi} sin \theta \ket{111} $

  The density operator can be expressed as $\rho= \begin{bmatrix}
     cos^2\theta& 0 & 0 & 0 & 0 & 0 & 0 & e^{i\phi}cos\theta sin\theta  \\  0 & 0 & 0 & 0 & 0 & 0 & 0 & 0 \\  0 & 0 & 0 & 0 & 0 & 0 & 0 & 0 \\  0 & 0 & 0 & 0 & 0 & 0 & 0 & 0 \\ 0 & 0 & 0 & 0 & 0 & 0 & 0 & 0 \\ 0 & 0 & 0 & 0 & 0 & 0 & 0 & 0 \\ 0 & 0 & 0 & 0 & 0 & 0 & 0 & 0 \\ e^{-i\phi}cos\theta sin\theta & 0 & 0 & 0 & 0 & 0 & 0 & sin^2\theta 
    \end{bmatrix}$
    $$= \begin{bmatrix} A^{11} & A^{12} \\ A^{21} & A^{22} \end{bmatrix}$$
    
   \item  $W= \frac{1}{\sqrt{3}} [\ket{001}+\ket{010}+\ket{100}$
   
 The density operator can be expressed as $ \rho= \frac{1}{3}\begin{bmatrix}
   0 & 0 & 0 & 0 & 0 & 0 & 0 & 0 \\  0 & 1 & 1 & 0 & 1 & 0 & 0 & 0 \\  0 & 1 & 1 & 0 & 1 & 0 & 0 & 0 \\  0 & 0 & 0 & 0 & 0 & 0 & 0 & 0 \\ 0 & 1 & 1 & 0 & 1 & 0 & 0 & 0 \\ 0 & 0 & 0 & 0 & 0 & 0 & 0 & 0 \\ 0 & 0 & 0 & 0 & 0 & 0 & 0 & 0 \\ 0 & 0 & 0 & 0 & 0 & 0 & 0 & 0
    \end{bmatrix} $
    $$=\begin{bmatrix} A^{11} & A^{12} & A^{13}  &  A^{14} \\ A^{21} & A^{22} & A^{23}  &  A^{24} \\ A^{31} & A^{32} & A^{33}  &  A^{34} \\A^{41} & A^{42} & A^{43}  &  A^{44}  \end{bmatrix}$$
    
   \item $W$-type state = $\frac{1}{\sqrt{2+2n}} [\ket{100}+ \sqrt{n} e^{i\delta} \ket{010} +\sqrt{n+1} e^{i\delta} \ket{001}$
   
    The density operator can be expressed as 
    $$\rho=\frac{1}{2(n+1)}  \begin{bmatrix}
    0 & 0 & 0 & 0 & 0 & 0 & 0 & 0 \\
    0 & n+1 & \sqrt{n(n+1)} & 0 & e^{i\delta} \sqrt{n+1}  & 0 & 0 & 0 \\ 
    0 & \sqrt{n(n+1)}  & n & 0 &  e^{i\delta} \sqrt{n} & 0 & 0 & 0 \\
    0 & 0 & 0 & 0 & 0 & 0 & 0 & 0 \\
     0 &  e^{-i\delta} \sqrt{n+1} & e^{-i\delta}  \sqrt{n}  & 0 & 1 & 0 & 0 & 0 \\
   0 & 0 & 0 & 0 & 0 & 0 & 0 & 0 \\
   0 & 0 & 0 & 0 & 0 & 0 & 0 & 0 \\
   0 & 0 & 0 & 0 & 0 & 0 & 0 & 0 
    \end{bmatrix} $$
    
   
    $$ =\begin{bmatrix} A^{11} & A^{12} & A^{13}  &  A^{14} \\ A^{21} & A^{22} & A^{23}  &  A^{24} \\ A^{31} & A^{32} & A^{33}  &  A^{34} \\A^{41} & A^{42} & A^{43}  &  A^{44}  \end{bmatrix}$$
\end{enumerate}
     We can clearly see that blocks of all entangled states considered here cannot be decomposed as $A^{ij}$ as $A^{ij}=B^{ij} - C^{ij}+ iD^{ij}-iE^{ij} $ for all $i \neq j$  and $A^{ii} - B^{ij} - C^{ij} - D^{ij}-E^{ij} \ngeq 0$, therefore, all considered states are entangled.
\end{enumerate}

\bibliographystyle{plain}
\bibliography{Dogu}

\end{document}